\documentclass[aps,prx,twocolumn,superscriptaddress]{revtex4-2}
\usepackage{bm}
\usepackage{graphicx}
\usepackage{color}
\usepackage{amsmath}
\usepackage{amssymb}
\usepackage[colorlinks=true,linkcolor=blue,anchorcolor=red,citecolor=blue,urlcolor=blue]{hyperref}

\makeatletter

\def\Fig#1{\text{Fig.}~\ref{#1}}

\DeclareUnicodeCharacter{2212}{-}

\begin{document}

\title{Nearly Isotropic Vortex Solid in $\mathbf{(La,Pr)_{3}Ni_{2}O_{7}}$ Thin Films}

\author{Yaolong Bian}
\thanks{These authors contributed equally to this work.}
\affiliation{Anhui Key Laboratory of Low-Energy Quantum Materials and Devices, High Magnetic Field Laboratory (CHMFL), Hefei Institutes of Physical Science, Chinese Academy of Sciences, Hefei 230031, China}
\affiliation{Science Island Branch, Graduate School of USTC, Hefei 230026, China}

\author{Yaqi Chen}
\thanks{These authors contributed equally to this work.}
\affiliation{State Key Laboratory of Quantum Functional Materials, Department of Physics, Guangdong Provincial Key Laboratory of Topological Matter, Guangdong Basic Research Center of Excellence for Quantum Science, and College of Semiconductors (National Graduate College for Engineers), Southern University of Science and Technology, Shenzhen, China}
\affiliation{UWBG \textsf{\&} UNBG Materials and Power Devices Research Group, Shenzhen Pinghu Laboratory, Shenzhen 518111, China}

\author{Heng Wang}
\thanks{This author contributed equally to this work}
\thanks{Contact author: wangheng@quantumsc.cn} 
\affiliation{State Key Laboratory of Quantum Functional Materials, Department of Physics, Guangdong Provincial Key Laboratory of Topological Matter, Guangdong Basic Research Center of Excellence for Quantum Science, and College of Semiconductors (National Graduate College for Engineers), Southern University of Science and Technology, Shenzhen, China}
\affiliation{Quantum Science Center of Guangdong-Hong Kong-Macao Greater Bay Area, Shenzhen 518045, China}

\author{Guangdi Zhou}
\affiliation{State Key Laboratory of Quantum Functional Materials, Department of Physics, Guangdong Provincial Key Laboratory of Topological Matter, Guangdong Basic Research Center of Excellence for Quantum Science, and College of Semiconductors (National Graduate College for Engineers), Southern University of Science and Technology, Shenzhen, China}
\affiliation{Quantum Science Center of Guangdong-Hong Kong-Macao Greater Bay Area, Shenzhen 518045, China}

\author{Fei Peng}
\affiliation{State Key Laboratory of Quantum Functional Materials, Department of Physics, Guangdong Provincial Key Laboratory of Topological Matter, Guangdong Basic Research Center of Excellence for Quantum Science, and College of Semiconductors (National Graduate College for Engineers), Southern University of Science and Technology, Shenzhen, China}

\author{Zichen Lv}
\affiliation{Anhui Key Laboratory of Low-Energy Quantum Materials and Devices, High Magnetic Field Laboratory (CHMFL), Hefei Institutes of Physical Science, Chinese Academy of Sciences, Hefei 230031, China}
\affiliation{Science Island Branch, Graduate School of USTC, Hefei 230026, China}

\author{Jiaqiang Cai}
\affiliation{Anhui Key Laboratory of Low-Energy Quantum Materials and Devices, High Magnetic Field Laboratory (CHMFL), Hefei Institutes of Physical Science, Chinese Academy of Sciences, Hefei 230031, China}
\affiliation{Science Island Branch, Graduate School of USTC, Hefei 230026, China}

\author{Yifan chen}
\affiliation{Anhui Key Laboratory of Low-Energy Quantum Materials and Devices, High Magnetic Field Laboratory (CHMFL), Hefei Institutes of Physical Science, Chinese Academy of Sciences, Hefei 230031, China}
\affiliation{Science Island Branch, Graduate School of USTC, Hefei 230026, China}

\author{Wenjie Meng}
\affiliation{Anhui Key Laboratory of Low-Energy Quantum Materials and Devices, High Magnetic Field Laboratory (CHMFL), Hefei Institutes of Physical Science, Chinese Academy of Sciences, Hefei 230031, China}

\author{Ze Wang}
\affiliation{Anhui Key Laboratory of Low-Energy Quantum Materials and Devices, High Magnetic Field Laboratory (CHMFL), Hefei Institutes of Physical Science, Chinese Academy of Sciences, Hefei 230031, China}

\author{Haoliang Huang}
\affiliation{State Key Laboratory of Quantum Functional Materials, Department of Physics, Guangdong Provincial Key Laboratory of Topological Matter, Guangdong Basic Research Center of Excellence for Quantum Science, and College of Semiconductors (National Graduate College for Engineers), Southern University of Science and Technology, Shenzhen, China}
\affiliation{Quantum Science Center of Guangdong-Hong Kong-Macao Greater Bay Area, Shenzhen 518045, China}

\author{Daohua Zhang}
\affiliation{State Key Laboratory of Quantum Functional Materials, Department of Physics, Guangdong Provincial Key Laboratory of Topological Matter, Guangdong Basic Research Center of Excellence for Quantum Science, and College of Semiconductors (National Graduate College for Engineers), Southern University of Science and Technology, Shenzhen, China}
\affiliation{UWBG \textsf{\&} UNBG Materials and Power Devices Research Group, Shenzhen Pinghu Laboratory, Shenzhen 518111, China}

\author{Mingliang Tian}
\affiliation{Anhui Key Laboratory of Low-Energy Quantum Materials and Devices, High Magnetic Field Laboratory (CHMFL), Hefei Institutes of Physical Science, Chinese Academy of Sciences, Hefei 230031, China}

\author{Jinfeng Jia}
\affiliation{State Key Laboratory of Quantum Functional Materials, Department of Physics, Guangdong Provincial Key Laboratory of Topological Matter, Guangdong Basic Research Center of Excellence for Quantum Science, and College of Semiconductors (National Graduate College for Engineers), Southern University of Science and Technology, Shenzhen, China}
\affiliation{Quantum Science Center of Guangdong-Hong Kong-Macao Greater Bay Area, Shenzhen 518045, China}

\author{Qi-kun Xue}
\affiliation{State Key Laboratory of Quantum Functional Materials, Department of Physics, Guangdong Provincial Key Laboratory of Topological Matter, Guangdong Basic Research Center of Excellence for Quantum Science, and College of Semiconductors (National Graduate College for Engineers), Southern University of Science and Technology, Shenzhen, China}
\affiliation{Quantum Science Center of Guangdong-Hong Kong-Macao Greater Bay Area, Shenzhen 518045, China}
\affiliation{Department of Physics, Tsinghua University, Beijing, China}

\author{Zhuoyu Chen}
\thanks{Contact author: chenzhuoyu@sustech.edu.cn} 
\affiliation{State Key Laboratory of Quantum Functional Materials, Department of Physics, Guangdong Provincial Key Laboratory of Topological Matter, Guangdong Basic Research Center of Excellence for Quantum Science, and College of Semiconductors (National Graduate College for Engineers), Southern University of Science and Technology, Shenzhen, China}
\affiliation{Quantum Science Center of Guangdong-Hong Kong-Macao Greater Bay Area, Shenzhen 518045, China}

\author{Jinglei Zhang}
\thanks{Contact author: zhangjinglei@hmfl.ac.cn} 
\affiliation{Anhui Key Laboratory of Low-Energy Quantum Materials and Devices, High Magnetic Field Laboratory (CHMFL), Hefei Institutes of Physical Science, Chinese Academy of Sciences, Hefei 230031, China}

\begin{abstract}
The discovery of superconductivity in bulk bilayer nickelates has established a new platform for exploring high-$T_c$ superconductivity beyond the cuprates. The role of the Ni $3d_{z^2}$-derived $\gamma$ band in the superconductivity of bilayer nickelates remains unresolved. By performing simultaneous resistance and diamagnetism measurements on (La,Pr)$_3$Ni$_2$O$_7$ thin films, we map the vortex melting phase diagram for both in-plane and out-of-plane magnetic fields. For $H\parallel c$, the geometric confinement effect gives rise to pancake vortices. 
Remarkably,  the anisotropy parameter of the vortex melting field  $\gamma_{H_m} \equiv H_m^{ab}/H_m^c$ decreases monotonically with decreasing temperature and approaches unity at low temperatures. Within the anisotropic Ginzburg--Landau scaling, $H_m^{ab}/H_m^c = \sqrt{\rho_s^{ab}/\rho_s^c}$ tracks the superfluid-density anisotropy. Such a vortex solid implies a nearly isotropic superfluid density, which is irreconcilable with the strictly two-dimensional $3d_{x^2-y^2}$-derived bands, but naturally explained by a substantial interlayer superfluid contribution from the $3d_{z^2}$-derived $\gamma$ band. Our results provide thermodynamic evidence for a substantial contribution of the $\gamma$ band to superconductivity in bilayer nickelate thin films.


\end{abstract}
	
\maketitle
	
\begin{figure*}[!htbp]
    \centering
	\includegraphics[width=17cm]{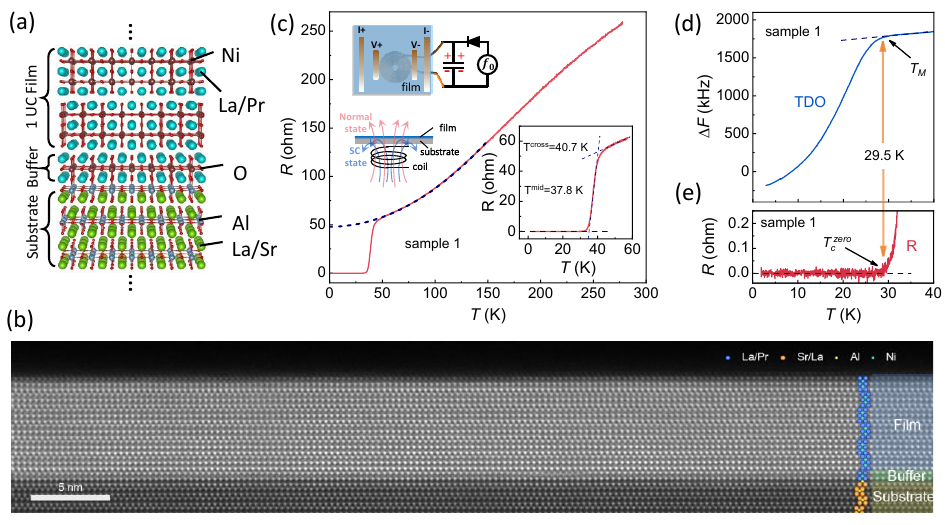}
	\caption{\textbf{Sample characterization of ambient-pressure bilayer nickelate $\mathbf{(La,Pr)_{3}Ni_{2}O_{7}}$.} 
    (a) Structural schematic of the $\mathrm{(La,Pr)}_{3}\mathrm{Ni}_{2}\mathrm{O}_{7}$ bilayer film, grown on a (001)-oriented SrLaAl$\mathrm{O}_{4}$ substrate. 
    (b) Cross-sectional high-angle annular dark-field scanning transmission electron microscopy (HAADF-STEM) image of a 3UC $\mathrm{(La,Pr)}_{3}\mathrm{Ni}_{2}\mathrm{O}_{7}$/SrLaAl$\mathrm{O}_{4}$ sample, with the size scale indicated in the lower left of 5 nm.
    (c) Temperature dependence of the in-plane resistance of the $\mathrm{(La,Pr)}_{3}\mathrm{Ni}_{2}\mathrm{O}_{7}$ bilayer film. The blue dashed line is a power-law fit  of the normal-state resistivity over the 50-150 K range. The upper inset shows the schematic of the TDO setup, which allows simultaneous measurement of the sample's resistance and diamagnetic response. The lower inset is a zoom-in view of the in-plane resistance near the superconducting transition.
    (d) and (e) represent the temperature dependence of the TDO frequency and in-plane resistance at low temperature, respectively. The yellow arrows are guides to the eye.}
    \label{fig.1}
	\end{figure*}

\emph{Introduction--} The discovery of superconductivity in bilayer La$_3$Ni$_2$O$_7$ and trilayer La$_4$Ni$_3$O$_{10}$ under high pressure has established Ruddlesden--Popper nickelates as a new platform for high-$T_c$ superconductors\cite{mWang2023n,hqYuan2024np,jgChen2024prx,hhWen2024cpl,jZhao2024n,ypQi2025prx,jgChen2024n}. Unlike the cuprates, where the Cu $3d_{x^2-y^2}$ orbital alone drives Cooper pairing\cite{zaanen2015n}, the bilayer nickelates feature a multiorbital low-energy manifold involving both $3d_{x^2-y^2}$ and $3d_{z^2}$ orbitals. This multiorbital character has led to competing theoretical scenarios for the pairing mechanism, with models ranging from interlayer $s_\pm$-wave pairing driven by $3d_{z^2}$ hybridization\cite{Luo2023Bilayer,PhysRevB.108.L140505, Sakakibara2024Possible, Gu2025Effective, Liu2023sWave, Qu2024Bilayer, Yang2023Interlayer, PhysRevB.108.165141, Huang2023Impurity, Qin2023HighTc, Tian2024Correlation,Zhang2024Prediction, Zhang2024Structuralphase, PhysRevB.110.L180501, Huo2025Modulation, Jiang2025Theory, Borchia2025Extended,10.1093/nsr/nwaf253} to $d$-wave pairing dominated by in-plane $3d_{x^2-y^2}$ interactions\cite{Lechermann2023Electronic,Jiang2024HighTemperature,Lu2024Interlayer,PhysRevB.110.205122,Oh2023TypeII}. The recent stabilization of ambient-pressure superconductivity in compressively strained thin films has enabled direct spectroscopic investigations\cite{zyChen2026sci,Hwang2026prx,yfNie2026np,zyChen2026prl,xhChen2026arpes}. However, even angle-resolved photoemission spectroscopy (ARPES) measurements alone have yielded conflicting results: some studies find the $\gamma$ band crossing the Fermi level with a sizable superconducting gap\cite{zyChen2026prl,zyChen2026sci,xhChen2026arpes}, while others place it below $E_F$ with no observable gap\cite{yfNie2026np,Hwang2026prx}.
Thus, the role of the $3d_{z^2}$ orbital---and hence the fundamental pairing mechanism---remains an open question.

The upper critical field $H_{c2}$, particularly its anisotropy, provides a powerful tool to probe the electronic structure. $H_{c2}$ of pressurized RP nickelates appears largely isotropic\cite{ypQi2025jacs,jlZhang2026prx}. In contrast, compressively strained thin films exhibit a two-dimensional (2D) behavior of $H_{c2}$ near $T_c$\cite{zyChen2025n,yfNie2025nm,Hwang2026nc,ywXie2026am,zxShi2026mt,xhChen2026arxiv,Tsukazaki2026nm}. Notably, the anisotropy of $H_{c2}$ in thin films is complicated by the strong Pauli paramagnetic effect, which competes with the orbital depairing channel and masks the intrinsic electronic structure. Alternatively, the diamagnetic response directly probes the equilibrium vortex state and provides a cleaner measure of the superconducting anisotropy\cite{RevModPhys.66.1125}. However, for RP nickelate thin films, the Meissner signal is severely suppressed by sample imperfections and the metastable nature of the superconducting phase, making such measurements challenging, especially at high magnetic fields. As a result, the characteristic fields extracted from diamagnetic responses in this system remain largely unexplored.


In this work, we establish a comprehensive vortex phase diagram for compressively strained $\mathrm{(La,Pr)_3Ni_2O_7}$ thin films through simultaneous measurements of electrical resistance and diamagnetism. For $\mathbf{H} \parallel c$, the geometric confinement of the ultrathin film gives rise to pancake vortex behavior in the vortex melting response. 
For samples with different vortex liquid regions, the anisotropy ratio $\gamma_{H_m} = H_m^{ab}/H_m^{c}$ exhibits a similar downward trend as temperature decreases, extrapolating towards unity as $T \to 0$, which indicates a nearly isotropic vortex solid in the low-temperature limit.
Since $H_m$ is governed by the superfluid density, such isotropy suggests an isotropic superfluid response, pointing to an additional contribution from the Ni-$3d_{z^2}$-derived $\gamma$ band that naturally provides interlayer superfluid weight.

\begin{figure*}[!tbp]
    \centering
    \includegraphics[width=17cm]{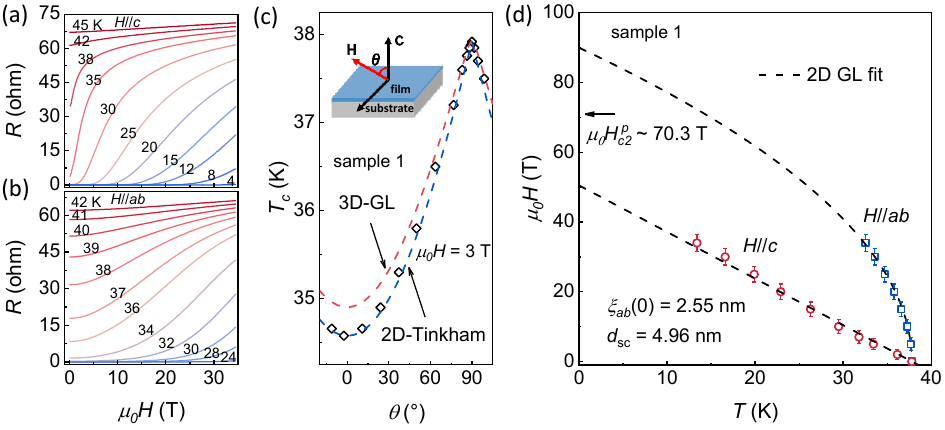}
    \caption{\textbf{Magnetic field responses of the resistance of the $\mathbf{(La,Pr)_{3}Ni_{2}O_{7}}$ bilayer film.} 
    (a), (b) Magnetic field dependence of resistance at selected temperatures with $\mathbf{H} \parallel c$ (a) and $\mathbf{H} \parallel ab$ (b). 
    (c) Angular dependence of the superconducting transition temperature $T_{c}$ under the magnetic field of 3 T, fitted with the 2D Tinkham model (blue dashed line) and the 3D GL model (red dashed line). Inset indicates the specific configuration during the angle-dependent measurements, where $\theta$ is the angle between magnetic field and the $c$ axis of the sample.
    (d) In-plane (open squares) and out-of-plane (open circles) critical fields, determined from the midpoint of the superconducting transitions, are plotted as functions of temperature. Black dashed lines are fits using the 2D-GL model. The error bars represent the uncertainty of magnetic field measurements.}
    \label{fig.2}
    \end{figure*}

\emph{Experimental details--} 
The $\mathrm{(La,Pr)}_{3}\mathrm{Ni}_{2}\mathrm{O}_{7}$ films (three-unit-cell-thick)  were grown on a (001)-oriented SrLaAlO$_4$ substrates by gigantic-oxidative atomic layer-by-layer epitaxy (\Fig{fig.1}(a)). Details of the growth procedure are provided in Ref.~\cite{zyChen2025n}.
The resistance is measured using the ac lock-in technique with electrodes on the top surface of the sample. The diamagnetic response is probed concurrently by a home-made tunnel diode oscillator (TDO) setup via a small disk-shaped coil closely attached to the bottom surface of the film (see the inset of \Fig{fig.1}(c)). Upon entering the Meissner state, the sample rearranges the coil's magnetic field distribution, thereby changing its inductance and shifting the TDO oscillation frequency. With this configuration, we can simultaneously access the two fundamental characteristics of superconductors---zero resistance and Meissner diamagnetism. Transport and diamagnetism measurements under high magnetic fields up to 34~T were carried out at the Chinese High Magnetic Field Laboratory in Hefei, using a resistive water-cooled magnet (WM5).

\emph{Sample characterization--} 
\Fig{fig.1}(b) presents the large-field-of-view HAADF image of the 3UC $\mathrm{(La,Pr)}_{3}\mathrm{Ni}_{2}\mathrm{O}_{7}$ /SrLaAl$\mathrm{O}_{4}$ thin film. The film exhibits uniform, well-defined interfaces over extended regions and shows no observable intergrowths of adjacent RP phases, confirming its phase purity.
\Fig{fig.1}(c) and (d) present the zero-field resistance and diamagnetism of our $\mathrm{(La,Pr)}_{3}\mathrm{Ni}_{2}\mathrm{O}_{7}$ thin film. The normal state exhibits metallic behavior over a wide temperature range, following the power-law relation $R(T)=R_0+AT^n$ with an exponent $n = 2.0$, contrasting with the $T$-linear resistivity reported for bulk samples under pressure\cite{mWang2023n,hqYuan2024np,jZhao2024n}. A clear superconducting transition is observed in resistance with a cross at 40.7 K and a midpoint at 37.8 K. Upon further cooling to 29.5 K, zero resistance is reached within the noise level of the measurement system. The Meissner effect is simultaneously probed by the TDO setup, where the deviation of $f_{\mathrm{TDO}}$ from the normal-state background defines the vortex melting temperature $T_M$ (\Fig{fig.1}(d)). Remarkably, $T_M$ of our sample is significantly higher than that in previous studies\cite{xhChen2026arxiv,zyChen2025n,Hwang2025nm,zhNie2026n} and nearly coincides with the zero-resistance temperature $T_c^{\rm zero}$ (\Fig{fig.1}(e)), which fully demonstrates the high quality and good superconducting homogeneity of our bilayer nickelate thin films.


\begin{figure*}[!tbp]
    \centering
	\includegraphics[width=14cm]{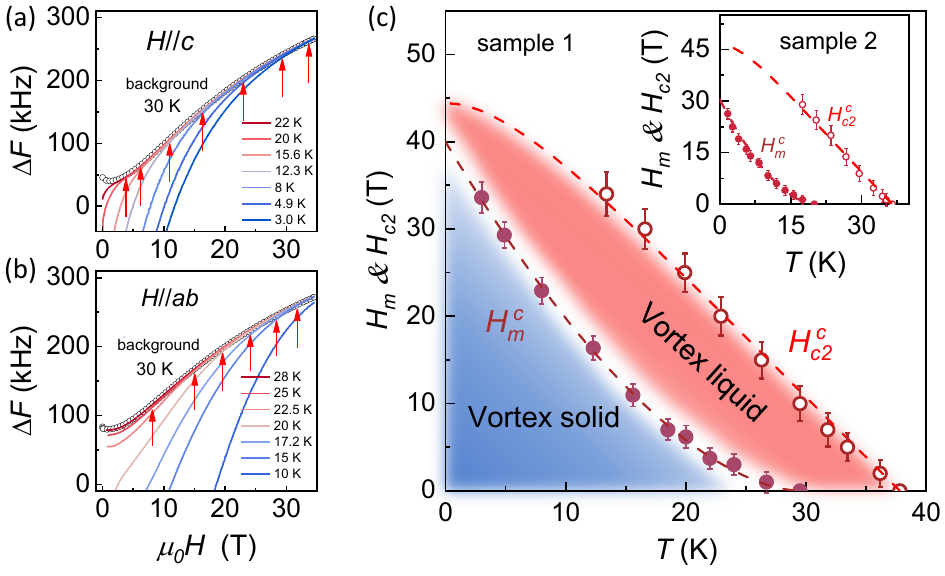}
	\caption{\textbf{Magnetic-field dependence of the TDO frequency shift of the $\mathbf{(La,Pr)_{3}Ni_{2}O_{7}}$ film.} 
    (a),(b) Magnetic-field dependence of TDO frequency shift $\Delta F$ at selected temperatures with $H\parallel c$ (a) and $H\parallel ab$ (b). Open circles are $\Delta F$ taken at T = 30 K as a normal-state background signal. The arrows indicate vortex melting fields ($H_m$) determined from the point deviating from the background signal. 
    (c) The \textit{T}-\textit{H} phase diagram for the $\mathrm{(La,Pr)}_{3}\mathrm{Ni}_{2}\mathrm{O}_{7}$ bilayer film inferred from $H_{c2}(T)$ and $H_m(T)$. The open circles and solid circles are taken from \Fig{fig.2}(a) and \Fig{fig.3}(a). The red dashed line is a fit using the two-band model (see Supplemental Material). The deep red dashed line is a fit using the function of $H_m$. The inset is the \textit{T}-\textit{H} phase diagram of sample 2. The error bars represent the uncertainty of magnetic field measurements.}
    \label{fig.3}
	\end{figure*}

\emph{Confined quasi-2D superconductivity near $T_c$--} In order to estimate the anisotropy of the superconductivity, the measurements were conducted with the magnetic field applied along the out-of-plane (\Fig{fig.2}(a)) and in-plane (\Fig{fig.2}(b)) directions. The superconducting transition gradually shifts to a lower magnetic field with increasing temperature. 
Fig. S3 shows a significantly broader transition width for $\mathbf{H} \parallel c$ than that for $\mathbf{H} \parallel ab$ as field strength increases, indicating that the nickelate films display a highly anisotropic response to different magnetic field orientations. As displayed in \Fig{fig.2}(c), the angular dependence of the superconducting transition temperature $T_c(\theta)$ near $T_c$ displays a sharp cusp-like feature that conforms well to the two-dimensional (2D) Tinkham model\cite{Tinkham1963Effect} while deviating markedly from the three-dimensional Ginzburg--Landau (GL) description\cite{PhysRevB.40.5263}, establishing the quasi-2D nature of the superconductivity. As shown in \Fig{fig.2}(d), the $H_{c2}(T)$ data for both $\mathbf{H}\parallel c$ and $\mathbf{H}\parallel ab$ can be nicely described by the 2D-GL model\cite{Wang2021Isotropic}, resulting in zero-temperature critical fields of 90.0 T for in-plane and 50.5 T for out-of-plane, respectively. Coherence lengths can be estimated using the relation $\xi_0 = ({\phi_0}/{2\pi H_{c2}})^{1/2}$\cite{zyChen2025n}, yielding in-plane and out-of-plane coherence lengths of $\xi_{ab} \sim 2.55$ nm and $\xi_c \sim 1.43$ nm. 
Meanwhile, effective superconducting thickness can be estimated using the formula $d_{\mathrm{SC}}=({6\phi_0 H_c}/{\pi H_{ab}^2})^{1/2}$,  yielding $d_{\mathrm{SC}}\sim 4.96$ nm, which is close to the film thickness of 6.8 nm, confirming the intrinsic bulk nature of the superconductivity in our film.
The comparable magnitudes of these length scales suggest that the bilayer nickelate thin-film superconductivity resides in a geometrically confined 2D regime. Further analysis shows orbital depairing and multiband effects yield a linear $H_{c2}^c(T)$, while Pauli paramagnetism dominates $H_{c2}^{ab}$ (see Supplemental Material Fig. S5). The observed $H_{c2}$ anisotropy is therefore governed by competing depairing channels rather than intrinsic electronic structure.

\emph{Pancake-shaped vortices --} To further investigate the thermodynamic behavior of the critical field, we measured the field-dependent diamagnetic response of the $\mathrm{(La,Pr)}_{3}\mathrm{Ni}_{2}\mathrm{O}_{7}$ bilayer film using the TDO setup, which has been described in detail above. 
\Fig{fig.3}(a) and (b)  show frequency shift ($\Delta F$) as a function of the magnetic field at selected temperatures for $\mathbf{H} \parallel c$ and $\mathbf{H} \parallel ab$, respectively. $\Delta F$ increases with magnetic field and eventually converges to the normal-state background, which is taken at $T$ = 30 K. The $H_{TDO}$ is determined from the point where $\Delta F$ intercepts the background (detailed information in Supplemental Material). First, we focus on the out-of-plane magnetic field direction. The $H_{TDO}$ differs significantly from $\mu_0 H_{c2}$ but agrees well with that determined by the 0.1$\%$ resistance criterion (see Supplemental Material Fig. S6). Moreover, our previous mutual-inductance measurements on the $\mathrm{(La,Pr)}_{3}\mathrm{Ni}_{2}\mathrm{O}_{7}$ bilayer film have revealed that the onset of diamagnetism is associated with a sudden dissipative process due to vortex depinning\cite{zyChen2026nsr}. Thus, consistent with this dissipative nature, $H_{TDO}$ is identified with the vortex melting field $H_m$, which separates the vortex solid (blue) below it from the vortex liquid (red) between $H_m(T)$ and $H_{c2}(T)$ in \Fig{fig.3}(c).

The melting line $H_m^{c}$ can be described by the empirical formula\cite{PhysRevB.58.8826,Hussey1999Melting} 
\begin{equation}
	H_m=H_0(1-T/T_M)^n,
\end{equation} where $T_M$ is the vortex melting temperature at zero magnetic field defined by \Fig{fig.1}(d).
The fitting yields an exponent $n^c = 1.7$, which falls in the characteristic range of 1.5--2.0 for pancake vortices in cuprates\cite{PhysRevLett.64.1063,RevModPhys.66.1125, PhysRevLett.67.2737,Hussey1999Melting}. Crucially, the exponent $n$ remains constant across the entire temperature range, suggesting that the pancake vortex morphology persists throughout the superconducting state. In the \textit{T}-\textit{H} phase diagram of $\mathrm{(La,Pr)}_{3}\mathrm{Ni}_{2}\mathrm{O}_{7}$ film, the vortex liquid region shrinks rapidly with decreasing temperature, with $H_m^c(T)$ approaching $H_{c2}^{c}$ at low temperatures. However, for another sample with a broader superconducting transition width than that of sample 1 (see Supplemental Material Fig. S8), the vortex liquid region remains extensive over the entire temperature range, indicating that disorder suppresses the vortex solid phase and broadens the liquid regime.

\begin{figure}[!tbp]
    \centering
    \includegraphics[width=\linewidth]{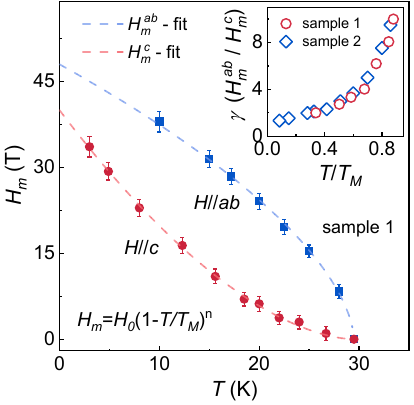}
    \caption{\textbf{The vortex melting fields of the $\mathbf{(La,Pr)_{3}Ni_{2}O_{7}}$ film.} 
    In-plane (solid squares) and out-of-plane (solid circles) $H_m$ are plotted as functions of temperature, which are taken from \Fig{fig.3}(b) and \Fig{fig.3}(a), respectively. Red and blue dashed lines are fits using a function of $H_{m}$. The error bars represent the uncertainty of magnetic field measurements. The inset shows the anisotropy of the melting fields for sample 1 and sample 2.}
    \label{fig.4}
	\end{figure}

\emph{Intrinsic isotropy of the vortex solid --} We now extend our discussion to the anisotropy of the melting field. The temperature dependence of $H_m^{ab}$ (solid squares) is well described by Eq.~(1), yielding $n^{ab}=0.6$, significantly lower than $n^c$. 
The smaller value of $n^{ab}$ is a hallmark of Josephson-vortex behavior, where vortices are elongated along the layers and their melting is governed by the much weaker interlayer Josephson coupling \cite{Hardy2020VortexMelting}. The anisotropy parameter of the vortex melting field, defined as $\gamma_{H_m} = H_m^{ab}/H_m^c$, decreases monotonically with decreasing temperature, from $\gamma_{H_m} \sim 10$ near $T_M$ to an extrapolated value of $\sim1.1$$(\pm0.1)$ in the zero-temperature limit (as shown in the inset of \Fig{fig.4}). This behavior can be understood by considering the effects of thermal fluctuations, which are known to be pronounced in such thin film systems due to their short coherence length and high anisotropy. Such effect substantially enhance the 2D behavior by weakening the interlayer Josephson coupling. As a result, the system exhibits enhanced anisotropic behavior near $T_M$. As temperature decreases, these thermal fluctuations are gradually quenched, and the intrinsic, nearly isotropic superfluid response of the system is revealed. 
Notably, the same near-isotropic behavior at low temperatures is also observed in samples with broader vortex liquid regions (see Supplemental Material Fig. S12), confirming that this isotropy is a generic feature of the bilayer nickelate system.

To interpret this isotropic vortex solid, it is useful to recall that $H_m$ and $H_{c2}$ weigh the condensate differently: $H_{c2}$ is the depairing scale at which the pair amplitude is suppressed, whereas $H_m$ is set by the rigidity of the vortex lattice --- a ground-state phase-stiffness scale. Within anisotropic Ginzburg--Landau theory\cite{Blatter1992From, RevModPhys.66.1125}, the Lindemann criterion --- with $H_m^c$ probing the in-plane shear stiffness and $H_m^{ab}$ reflecting the interlayer (Josephson) coupling --- yields $H_m^{ab}/H_m^c = \lambda_c/\lambda_{ab} = \sqrt{\rho_s^{ab}/\rho_s^c}$, with $\rho_s \propto \lambda^{-2}$. Therefore, the anisotropy of $H_m$ reflects, to a large extent, the anisotropy of the superfluid density. The convergence of $\gamma_{H_m}$ to unity at low temperatures thus implies $\rho_s^c \to \rho_s^{ab}$, i.e. an isotropic superfluid response. 

\emph{Discussion--}The role of the Ni-$3d_{z^2}$-derived $\gamma$ band in the superconductivity of RP nickelates remains debated. 
Our vortex melting anisotropy measurements offer a thermodynamic perspective on this issue. The $\alpha$ and $\beta$ bands, primarily of Ni $3d_{x^2-y^2}$ character, are confined to the NiO$_2$ planes and their superfluid contribution is expected to be predominantly in-plane. Even with fully gapped superconducting gaps, their orbital character limits their ability to provide substantial interlayer superfluid weight. The nearly isotropic vortex solid we observe thus points to an additional contribution. In the bilayer nickelate structure, the Ni-$3d_{z^2}$-derived $\gamma$ band is a natural candidate, as it provides interlayer coupling via the apical oxygen atoms\cite{Gu2025Effective, Liu2023sWave, Yang2023Interlayer}. Crucially, our ARPES measurements on the same films resolve a superconducting gap on the $\gamma$ band \cite{zyChen2026prl,zyChen2026sci}. A natural interpretation is that the $\gamma$-band superfluid provides the necessary interlayer superfluid weight, implying its direct participation in the superconducting pairing in our samples. 
We note, however, that our data do not rule out additional contributions from hybridization between the $\gamma$ and $\alpha/\beta$ bands, especially given that some ARPES studies place the $\gamma$ band below $E_F$. Quantifying the relative contributions will require future comparative studies on samples with controlled doping or growth conditions. Nonetheless, our present data establish the $\gamma$ band as a non-negligible component of the superconducting state and provide essential constraints for theoretical models.

\emph{Acknowledgments--}This work is supported by the Guangdong Major Project of Basic Research (No. 2025B0303000004), the National Key Research and Development Program of China (2022YFA1403101, 2024YFA1408101), the Natural Science Foundation of China (92565303, 92265112, 12374455, 12504161, 52672330, 12504166, \& 52388201), the Guangdong Project (2025TQ09A623), the Guangdong Provincial Quantum Science Strategic Initiative (GDZX2501001, GDZX2401004 \& GDZX2201001), the Shenzhen Science and Technology Program (KQTD20240729102026004), and the Shenzhen Municipal Funding Co-Construction Program Project (SZZX2301004 \& SZZX2401001), and the support from the Station of Quantum Materials.
This work at High Magnetic Field Laboratory was supported by the National Key R\&D Program of the MOST of China (Grants No. 2022YFA1602602, No. 2024YFA0727900), the National Natural Science Foundation of China (Grants No. 12474053), the Basic Research Program of the Chinese Academy of Sciences Based on Major Scientific Infrastructures (Grant No. JZHKYPT-2021-08). We thank the WM5\cite{wm5} at the Steady High Magnetic Field Facility, CAS\cite{CAS}, for providing technical support and assistance in data collection and analysis.


\bibliographystyle{apsrev4-1-etal-title_6authors}
\bibliography{LNO327}

\end{document}